\documentclass[10pt,twocolumn,letterpaper]{article}

\usepackage[T1]{fontenc}
\usepackage[utf8]{inputenc}
\usepackage{times}
\usepackage{geometry}
\usepackage{graphicx}
\usepackage{booktabs}
\usepackage{tabularx}
\usepackage{amsmath,amssymb}
\usepackage{caption}
\usepackage{microtype}


\newenvironment{promptquote}%
  {\begin{list}{}{%
     \setlength{\leftmargin}{1.2em}\setlength{\rightmargin}{0pt}%
     \setlength{\listparindent}{0pt}\setlength{\itemindent}{0pt}%
     \setlength{\topsep}{4pt}\setlength{\parsep}{3pt plus 1pt}%
     \setlength{\itemsep}{0pt}}\item[]\small}%
  {\end{list}}
\usepackage{xurl}
\usepackage[hidelinks]{hyperref}
\hypersetup{
  colorlinks=true,
  linkcolor=[rgb]{0.10,0.25,0.55},   
  citecolor=[rgb]{0.10,0.40,0.25},   
  urlcolor=[rgb]{0.10,0.25,0.75},
  breaklinks=true,
  pdftitle={UQMIA: An Open, Hands-On Tutorial on Uncertainty Quantification in Medical Imaging Analysis},
  pdfauthor={Gheiji, Elyassirad, Vatanparast, Tavakoli, Faghani}
}

\makeatletter
\renewcommand\@seccntformat[1]{\csname the#1\endcsname.\quad}
\makeatother

\begin{document}
\twocolumn[{%
\begin{center}
  {\LARGE\bfseries UQMIA: An Open, Hands-On Tutorial on Uncertainty Quantification in Medical Imaging Analysis with Large Language Model-Based Assessment of Educational Content \par}
  \vspace{1.1em}
  {\large Benyamin Gheiji\textsuperscript{1}, Danial Elyassirad\textsuperscript{1}, Mahsa Vatanparast\textsuperscript{2}, Meysam Tavakoli\textsuperscript{3}, Shahriar Faghani\textsuperscript{4,5*} \par}
  \vspace{0.7em}
  {\small
  1- Student Research Committee, Faculty of Medicine, Mashhad University of Medical Sciences, Mashhad, Iran  \\
  2- Transplant Research Center, Clinical Research Institute, Mashhad University of Medical Sciences, Mashhad, Iran  \\
  3- Department of Radiation Oncology, Northside Hospital, Atlanta GA  \\
  4- Radiology Informatics Lab, Department of Radiology, Mayo Clinic, Rochester, MN, USA  \\
  5- Department of Radiology, University of Pennsylvania, Philadelphia, PA, USA  \\
  (*) Correspondence: Shahriar Faghani, Email: Faghani.Shahriar@mayo.edu
  \par}
\end{center}
\vspace{1.4em}
}]
\begin{abstract}
\small
Uncertainty quantification (UQ) is increasingly recognized as an important component of reliable machine learning in medical imaging, yet practical resources connecting UQ theory, implementation, and evaluation remain limited. We developed Uncertainty Quantification in Medical Imaging Analysis (UQMIA), an open-access, hands-on tutorial comprising 19 sessions covering major UQ approaches, including variational inference, Monte Carlo dropout, deep ensembles, evidential deep learning, and conformal prediction, together with methods for evaluating uncertainty reliability. The tutorial is designed for researchers and practitioners with experience in medical imaging, progressing from foundational concepts to implementation and evaluation, with notebooks executable through Kaggle. Beyond presenting the tutorial, we introduce a framework using large language models (LLMs) to evaluate whether technical educational resources contain retrievable and usable knowledge. Using 100 four-option multiple-choice questions derived from primary methodological literature, we evaluated 20 instruction-tuned LLMs from six model families with and without retrieved tutorial context. Retrieval of UQMIA improved accuracy in 18 of 20 models, increasing mean accuracy from 0.680 to 0.742 (+0.062; Holm-adjusted P=0.00032), and improved area under the receiver operating characteristic curve (AUROC) in 18 of 20 models, increasing from 0.725 to 0.788 (+0.063; Holm-adjusted P=0.00032). UQMIA provides an accessible resource for learning UQ in medical imaging and demonstrates an LLM-based approach for evaluating technical educational material. The tutorial is available at \url{https://benyamin-gheiji.github.io/Uncertainty-Quantification-Medical-Imaging-Analysis/} and its source code and notebooks at \url{https://github.com/benyamin-gheiji/Uncertainty-Quantification-Medical-Imaging-Analysis/}.

\end{abstract}

\section{Introduction}

Clinical machine learning (ML) models are increasingly being developed for medical image interpretation, with performance commonly summarized using accuracy, area under the receiver operating characteristic curve (AUROC), sensitivity, and specificity. Although these metrics are essential for characterizing population-level performance, they provide limited information about the reliability of an individual prediction. A model may achieve high average accuracy while still producing incorrect predictions for particular patients, including cases that differ substantially from those encountered during training. Uncertainty quantification (UQ) has therefore been recognized as an important component of reliable medical artificial intelligence (AI), particularly when model predictions are intended to support clinical decision making \cite{r1,r2}. In radiology, where AI systems may encounter substantial variation in image quality, acquisition protocols, anatomy, and disease presentation, knowing when a prediction should be trusted can be as important as the prediction itself \cite{r3,r4}. Recent reviews have consequently identified UQ as an important direction for improving the reliability, interpretability, and clinical acceptability of medical imaging AI \cite{r5,r6,r7}.

The distinction between a correct prediction and a trustworthy prediction is particularly important because an appropriately uncertain prediction can prompt a second read, additional imaging, or confirmatory testing, whereas a confidently incorrect prediction may provide no indication that further assessment is warranted and may consequently propagate into subsequent clinical decisions. This concern has motivated substantial work on UQ and communicating model reliability to end users \cite{r2,r5}. Large-scale evaluations have further shown that UQ can deteriorate when the input distribution differs from the training distribution, highlighting the importance of evaluating reliability beyond conventional in-distribution predictive performance \cite{r8}.

The problem is compounded by the behavior of modern neural networks. Standard classification models produce normalized softmax scores for every input, including inputs that may be substantially different from the training data. A high softmax score therefore does not necessarily indicate that a prediction is reliable or that the model has encountered a familiar case \cite{r9}. Moreover, commonly used training loss functions such as cross-entropy optimize predictive accuracy by encouraging the correct class to receive a higher score, but they do not directly enforce that predicted probabilities correspond to the empirical likelihood of being correct. As a result, modern neural networks can be substantially miscalibrated, with predicted confidence failing to match observed accuracy \cite{r10}. These observations motivate a broader perspective on model performance in which predictive accuracy should be considered alongside the reliability of model confidence.

UQ encompasses several methodological families for addressing this problem. Bayesian neural networks  provide a probabilistic framework for representing uncertainty in model parameters \cite{r11}, while variational inference (VI) offers a tractable approximation to Bayesian posterior inference \cite{r12}. Monte Carlo (MC) dropout provides a practical approximation to Bayesian inference by retaining dropout during inference and using multiple stochastic forward passes to characterize predictive variability \cite{r13}. Deep ensembles provide an alternative non-Bayesian approach in which independently trained neural networks are combined to obtain predictive distributions and have demonstrated strong performance for UQ and robustness to distribution shift \cite{r14}. Evidential deep learning (EDL) instead models uncertainty directly by predicting the parameters of a Dirichlet distribution over class probabilities \cite{r15}. Conformal prediction (CP) takes a fundamentally different approach, constructing prediction sets or intervals with finite-sample coverage guarantees under exchangeability assumptions and without requiring a particular probabilistic model \cite{r16,r17}. Recent work has explored UQ across diverse medical imaging applications including classification, segmentation, registration, and other imaging tasks \cite{r18,r19,r20,r21}.

Performing UQ, however, does not guarantee that the resulting uncertainty estimate is useful. Calibration assesses the agreement between predicted probabilities and the observed frequency of correct predictions. A model is well calibrated when predictions assigned a probability of 0.8 are correct approximately 80\% of the time. It is an important component of evaluating the reliability of predictive confidence \cite{r10}. Selective prediction and risk--coverage analysis provide a framework for allowing a model to abstain from uncertain cases and refer them for human assessment \cite{r22}. Out-of-distribution (OOD) detection addresses whether an input differs from the distribution represented by the training data, while evaluation under dataset shift examines whether UQ remains informative when the conditions encountered at deployment differ from those during development \cite{r8,r23}. These considerations are particularly important in medical imaging, where UQ may otherwise appear useful under controlled test conditions while failing to identify unreliable predictions in clinically relevant settings. Recent medical imaging reviews have therefore emphasized that appropriate evaluation of UQ is as important as the choice of UQ method itself \cite{r1,r5,r6,r7}.

Despite the breadth of this literature, UQ remains a challenging topic for researchers and clinicians entering medical AI. The field spans Bayesian inference, probabilistic modeling, deep learning (DL), calibration, distribution shift, statistical prediction, and decision-making. Several reviews have provided comprehensive taxonomies of UQ methods and their applications in medical imaging \cite{r1,r5,r6,r7}. However, reviews primarily serve to organize and synthesize existing research; they do not necessarily provide a structured, hands-on path through the subject for a learner who wants to understand the concepts and implement the methods. This gap is particularly relevant for medical imaging researchers, who may be experienced in developing predictive models but have limited exposure to the theoretical and practical aspects of UQ. An accessible resource that connects clinical motivation with conceptual explanations, executable implementations, and systematic evaluation could therefore complement the existing methodological literature.

To address this educational gap, we developed Uncertainty Quantification in Medical Imaging Analysis (UQMIA), an open-access, hands-on tutorial comprising 19 sessions. The tutorial progresses from the motivation and foundations of UQ through major UQ approaches, including Bayesian perspectives, VI, MC dropout, deep ensembles, EDL, and CP, and subsequently focuses on evaluating whether UQ estimates are informative and clinically meaningful. The material is organized to connect conceptual explanations with executable implementations, with practical examples using medical imaging data.

A further question is how such an educational resource can be evaluated at scale. Traditional assessment of tutorials and educational materials typically relies on expert review, learner feedback, or studies involving human participants. Large language models (LLMs) provide a complementary opportunity to examine whether a resource can function as a retrievable source of domain-specific knowledge. Retrieval-augmented generation (RAG) combines a language model's parametric knowledge with externally retrieved information, allowing the model to condition its response on a supplied knowledge source \cite{r24}. In the present study, we used the tutorial as retrieved context and evaluated 20 LLMs on 100 four-option multiple-choice questions covering UQ in medical imaging. We quantified performance using accuracy and AUROC, allowing us to assess both the models' ability to identify the correct answer and their ability to discriminate between correct and incorrect responses using their reported confidence.

Accordingly, this work presents both an educational resource and an empirical framework for examining its use as an external knowledge source. We provide an open, structured tutorial covering major families of UQ methods for medical imaging, with conceptual material paired with practical implementations to provide a continuous path from clinical motivation to working code. The tutorial further emphasizes that producing a UQ estimate is only the beginning. Subsequent sessions focus on evaluating whether the estimate is informative through calibration, selective prediction, and other reliability-oriented analyses. Finally, we evaluate the tutorial as a retrievable knowledge resource across 20 LLMs using a standardized set of 100 multiple-choice questions and assess performance using both accuracy and AUROC. Together, these components provide an openly accessible learning resource for researchers and practitioners interested in UQ for medical imaging while demonstrating a scalable approach for quantitatively evaluating how contemporary language models use domain-specific educational material.

\section{Methods}

\begin{table*}[t]
\centering
\small
\caption{Structure of the tutorial}
\label{tab:t1}
\setlength{\tabcolsep}{4pt}
\begin{tabularx}{\textwidth}{@{}llX@{}}
\toprule
\textbf{Part} & \textbf{Sessions} & \textbf{Focus} \\
\midrule
Part 1 --- Foundations & 1--3 & Why accuracy alone is insufficient in clinical AI; aleatoric versus epistemic uncertainty; how clinicians already reason probabilistically. \\
Part 2 --- Core UQ Methods & 4--15 & The Bayesian perspective; VI; MC dropout; Deep Ensembles; EDL; CP. Each concept session is paired with a PyTorch implementation, and the part closes with a side-by-side comparison. \\
Part 3 --- Evaluation and Reliability & 16--18 & Calibration and reliability diagrams; risk--coverage analysis and selective prediction; OOD detection. All three combine concept and implementation in one notebook. \\
Part 4 --- Future Directions & 19 & Choosing a method for a clinical problem; open challenges; directions for the field. \\
\bottomrule
\end{tabularx}
\end{table*}

\begin{table*}[t]
\centering
\small
\caption{LLMs evaluated. All models are instruction-tuned variants.}
\label{tab:t2}
\setlength{\tabcolsep}{4pt}
\begin{tabularx}{\textwidth}{@{}lXl@{}}
\toprule
\textbf{Family} & \textbf{Models} & \textbf{Reference} \\
\midrule
Qwen2.5 & \href{https://huggingface.co/Qwen/Qwen2.5-0.5B-Instruct}{0.5B}, \href{https://huggingface.co/Qwen/Qwen2.5-1.5B-Instruct}{1.5B}, \href{https://huggingface.co/Qwen/Qwen2.5-3B-Instruct}{3B}, \href{https://huggingface.co/Qwen/Qwen2.5-7B-Instruct}{7B}, \href{https://huggingface.co/Qwen/Qwen2.5-14B-Instruct}{14B} & \cite{r30,r31} \\
Falcon3 & \href{https://huggingface.co/tiiuae/Falcon3-1B-Instruct}{1B}, \href{https://huggingface.co/tiiuae/Falcon3-3B-Instruct}{3B}, \href{https://huggingface.co/tiiuae/Falcon3-7B-Instruct}{7B}, \href{https://huggingface.co/tiiuae/Falcon3-10B-Instruct}{10B} & \cite{r32} \\
Phi & \href{https://huggingface.co/microsoft/Phi-3-mini-4k-instruct}{Phi-3-mini (3.8B)}, \href{https://huggingface.co/microsoft/Phi-3.5-mini-instruct}{Phi-3.5-mini (3.8B)}, \href{https://huggingface.co/microsoft/Phi-4-mini-instruct}{Phi-4-mini (3.8B)}, \href{https://huggingface.co/microsoft/Phi-3-medium-4k-instruct}{Phi-3-medium (14B)} & \cite{r33,r34} \\
Mistral & \href{https://huggingface.co/mistralai/Mistral-7B-Instruct-v0.3}{Mistral-7B-v0.3}, \href{https://huggingface.co/mistralai/Ministral-8B-Instruct-2410}{Ministral-8B}, \href{https://huggingface.co/mistralai/Mistral-Nemo-Instruct-2407}{Mistral-Nemo (12B)} & \cite{r35,r36} \\
Granite 3.1 & \href{https://huggingface.co/ibm-granite/granite-3.1-2b-instruct}{2B}, \href{https://huggingface.co/ibm-granite/granite-3.1-8b-instruct}{8B} & \cite{r37} \\
Gemma 3 & \href{https://huggingface.co/google/gemma-3-1b-it}{1B}, \href{https://huggingface.co/google/gemma-3-4b-it}{4B} & \cite{r38} \\
\bottomrule
\end{tabularx}
\end{table*}

\subsection{Tutorial Scope and Intended Audience}

The tutorial is designed for researchers and practitioners with experience in developing medical imaging models who want to complement their predictions with reliable and interpretable measures of confidence. Readers are expected to be comfortable with Python, to understand the fundamentals of machine learning, to have working familiarity with DL, and to have prior experience handling medical imaging data.

\subsection{Tutorial Structure}

The tutorial consists of 19 sessions organized into four parts (Table~\ref{tab:t1}), progressing from the motivation and foundations of UQ to implementation and evaluation. Part 1 introduces the role of UQ in clinical AI and key uncertainty concepts. Part 2 covers major UQ methods, with each concept paired with a practical implementation. Part 3 covers methods for assessing the reliability and usefulness of uncertainty estimates and Part 4 discusses method selection, open challenges, and future directions.

\subsection{Implementation and Delivery}

All implementation sessions use the same publicly available pediatric chest radiograph dataset \cite{r25} and a DenseNet-121 backbone \cite{r26}, implemented in PyTorch \cite{r27}. Where computationally necessary, the backbone is frozen and only the classification head is trained. This reflects a deliberate design choice to ensure that every notebook can run to completion on the free GPU tier of a hosted notebook service, allowing the entire tutorial to remain accessible to readers without access to institutional computing resources.

The material is released in three forms. The Jupyter notebooks and their figures are available in a public repository at \url{https://github.com/benyamin-gheiji/Uncertainty-Quantification-Medical-Imaging-Analysis/}. Each session is additionally published as a Kaggle notebook, executable in the browser with the dataset already attached. A website at \url{https://benyamin-gheiji.github.io/Uncertainty-Quantification-Medical-Imaging-Analysis/} renders every session with navigation between them for readers who prefer to read rather than run.

\subsection{Evaluation Design}

The evaluation treats the tutorial as an external knowledge source and examines whether providing it to an LLM improves the model's performance on questions concerning UQ. It is intended as a content-coverage and knowledge-utilization assessment, evaluating whether the tutorial contains relevant information in a form that can be retrieved and applied by the model.

Two outcomes were evaluated. Accuracy measures whether the model selected the correct answer. AUROC measures how well the confidence assigned to the selected answer discriminates between the model's correct and incorrect responses, providing an assessment of its ability to identify when its own answers are likely to be correct. These measures capture complementary aspects of performance. A model may achieve higher accuracy while becoming less informative in its confidence estimates, and therefore accuracy alone does not establish whether its confidence can distinguish reliable predictions from errors. Evaluating both outcomes is consistent with the tutorial's broader emphasis on assessing not only predictive performance but also the reliability of model confidence.

\subsection{Question Bank}

One hundred four-option multiple-choice items were generated with Claude Opus 5 to assess knowledge of UQ in ML, with particular emphasis on medical imaging applications. The items were divided equally into four categories of 25 items each. Recall items assessed definitions, mechanisms, and the operation of specific methods. Conceptual items required integration of multiple concepts or explanation of observed behavior. Counterintuitive items were designed around common misconceptions, with the corresponding incorrect intuition represented among the distractors. Applied items addressed practical and clinical considerations, including threshold selection, abstention policies, external validation, acquisition shift, inference-time computational constraints, and communication of uncertainty to clinicians.

To minimize circularity, the questions were generated from the primary methodological literature rather than from the tutorial itself. The generation prompt explicitly identified the methodological works used as the knowledge base, including foundational studies on MC dropout, deep ensembles, Bayesian neural networks, EDL, calibration, CP, OOD detection, dataset shift, and selective prediction \cite{r3,r4,r5,r6,r7,r8,r9,r10,r11,r12,r13,r14}. The prompt also imposed predefined constraints on the uniqueness and plausibility of the correct answer, distractor construction, option length, surface-level cues, question specificity, coverage of the literature, and use of concrete medical imaging contexts.

The complete question-generation prompt was as follows:

\begin{promptquote}

Construct 100 multiple-choice items assessing knowledge of uncertainty quantification in machine learning, with particular emphasis on applications to medical imaging.

Source constraint. Items must be derived from the primary methodological literature rather than from any tutorial, course, or secondary teaching material. Ground the items in the claims made by the following works: Gal and Ghahramani (Monte Carlo dropout, 2016); Lakshminarayanan et al. (deep ensembles, 2017); Blundell et al. (weight uncertainty in neural networks, 2015); Kendall and Gal (aleatoric and epistemic uncertainty, 2017); Sensoy et al. (evidential deep learning, 2018); Guo et al. (calibration of modern neural networks, 2017); Angelopoulos and Bates (conformal prediction, introductory treatment); Vovk et al. (algorithmic learning in a random world); Hendrycks and Gimpel (out-of-distribution detection baseline, 2017); Ovadia et al. (predictive uncertainty under dataset shift, 2019); and Geifman and El-Yaniv (selective prediction).

Composition. Produce four categories of 25 items each:

recall (identifiers R01--R25): definitions, mechanisms, and the operation of specific methods.

conceptual (C01--C25): items requiring the integration of two or more ideas, or an explanation of why a described behavior arises.

counterintuitive (X01--X25): items for which a common intuition yields an incorrect answer. That incorrect intuition must be represented among the distractors.

applied (A01--A25): items concerning clinical deployment, including threshold selection, abstention policies, external validation, acquisition shift, inference-time compute constraints, and communication of uncertainty to clinicians.

\textbf{Item construction requirements:}

Each item shall present exactly four options, of which exactly one is unambiguously correct.

Distractors must be individually defensible as incorrect for an identifiable reason: a documented misconception, a conflated adjacent concept, or a true statement that does not answer the question posed. Implausible or filler options are not acceptable.

Option lengths within an item must fall within approximately $\pm$15\% of one another. Where balancing is required, expand the distractors rather than abbreviating the correct option, so that option length carries no signal as to which option is correct.

Options must not be distinguishable by surface features. Avoid hedging qualifiers, absolute quantifiers confined to incorrect options, and grammatical disagreement with the stem.

Each item must be self-contained, with no reference to figures, other items, or preceding options.

No two items may assess the same fact. The bank should achieve substantive coverage across the cited literature.

Stems should comprise one or two sentences and pose a specific question. Avoid generic constructions such as ``Which of the following is true?''

Where the subject matter permits, situate the item in a concrete imaging context, including chest radiography, computed tomography, magnetic resonance imaging, or histopathology, in preference to abstract phrasing.

Output specification. Return a single JSON array and no other content. Each element must contain a question identifier, category, stem, four options, and the index of the correct answer. The correct option must occupy index 0 and the answer field must equal 0 for every item, with presentation order randomized downstream.

Verification. Before returning the bank, confirm for each item that exactly one option is correct, that the remaining three are defensibly incorrect, and that the four option lengths fall within the stated tolerance. Report the number of items for which the correct option is the longest of the four; under adequate length balancing this figure should approximate 25.

Output length. The complete bank of 100 items must be returned in a single response. Placeholders, abbreviated items, and truncated output are not acceptable.

\end{promptquote}

The resulting 100-item question bank comprised 25 items in each of the four predefined categories and was subsequently used as the common evaluation set for all 20 LLMs.

\subsection{Retrieval}

The tutorial condition provided the material through retrieval rather than as full text, because the 19 notebooks exceed the context windows of the smaller models. Each notebook was converted to plain text while preserving the original Markdown and fenced code blocks, then segmented at Markdown headings and grouped into chunks of approximately 400 tokens. Oversized sections were further divided at sentence boundaries to ensure that no chunk began or ended mid-sentence. One structural unit of overlap was retained between consecutive chunks.

The resulting chunks were embedded using BAAI/bge-large-en-v1.5 \cite{r28} and indexed with FAISS \cite{r29} using inner-product search over L2-normalized embeddings. Because the embedding model uses asymmetric query and passage representations, its query instruction prefix was applied only to queries and not to the retrieved passages. For each question, the query consisted of the item stem concatenated with all four answer options, and the five highest-scoring chunks were retrieved. Retrieval was performed once and cached before model evaluation, ensuring that every model received identical retrieved context. Thus, differences in retrieval did not contribute to differences in model performance.

\subsection{Models and Scoring}

Twenty instruction-tuned LLMs across six families were evaluated (Table~\ref{tab:t2}). Each model answered all 100 items twice: once with the question alone, and once with the retrieved passages prepended. Option order was shuffled once per item with a fixed seed and held identical across models and conditions. The model's answer was determined from the logits at the first token position following the generation prompt. The maximum logit among the four answer-letter tokens was selected as the predicted option, and a softmax was applied across the four logits to obtain normalized probabilities. For AUROC analysis, the confidence score was defined as the probability assigned to the model's predicted option.

\subsection{Statistical Analysis}

The LLM was treated as the unit of analysis, with each model contributing one paired observation for each evaluation metric. Differences between conditions were assessed using the two-sided Wilcoxon signed-rank test. P-values were adjusted for multiple comparisons across the two metrics using the Holm method. A two-sided adjusted P-value < 0.05 was considered statistically significant.

\section{Results}

\begin{figure*}[t]
\centering
\includegraphics[width=\textwidth]{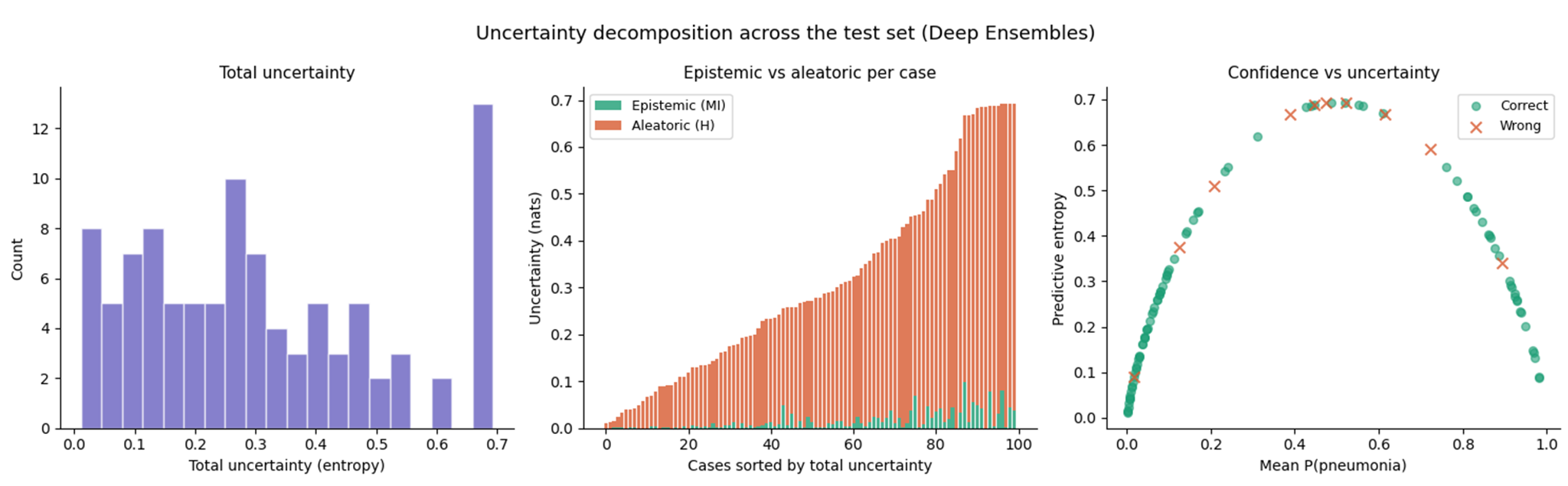}
\caption{Uncertainty decomposition produced by the Deep Ensembles implementation session, as described in the tutorial. From left to right, distribution of total predictive entropy across the test set; epistemic (mutual information) and aleatoric components per case, sorted by total uncertainty; and predictive entropy against mean predicted probability, with incorrect predictions marked. Errors concentrate in the high-entropy region.}
\label{fig:f1}
\end{figure*}

\begin{figure*}[t]
\centering
\includegraphics[width=\textwidth]{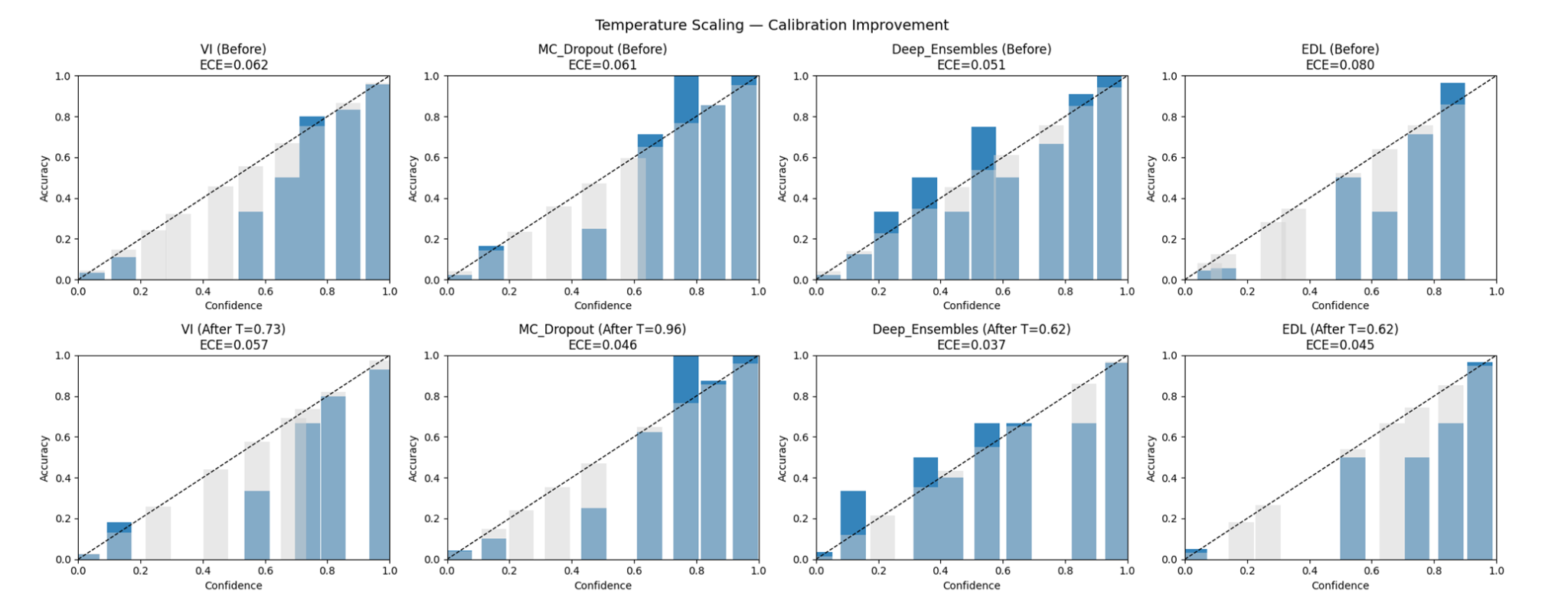}
\caption{Reliability diagrams for four methods before (top row) and after (bottom row) temperature scaling. ECE is reported for every method.}
\label{fig:f2}
\end{figure*}

\begin{figure*}[t]
\centering
\includegraphics[width=\textwidth]{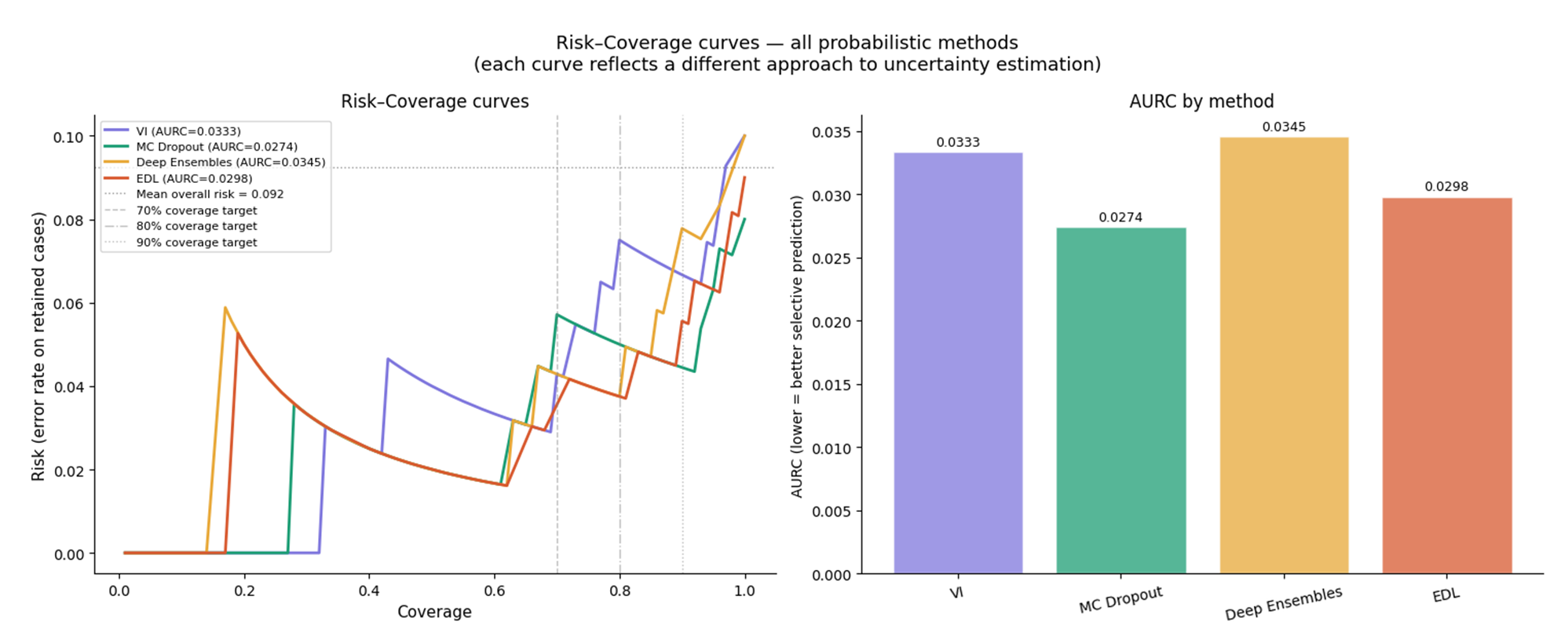}
\caption{Risk--coverage curves for the four probabilistic methods, with clinically motivated coverage targets marked. From left to right, risk against coverage; and area under the risk--coverage curve (AURC) by method, where lower is better.}
\label{fig:f3}
\end{figure*}

\begin{figure*}[t]
\centering
\includegraphics[width=\textwidth]{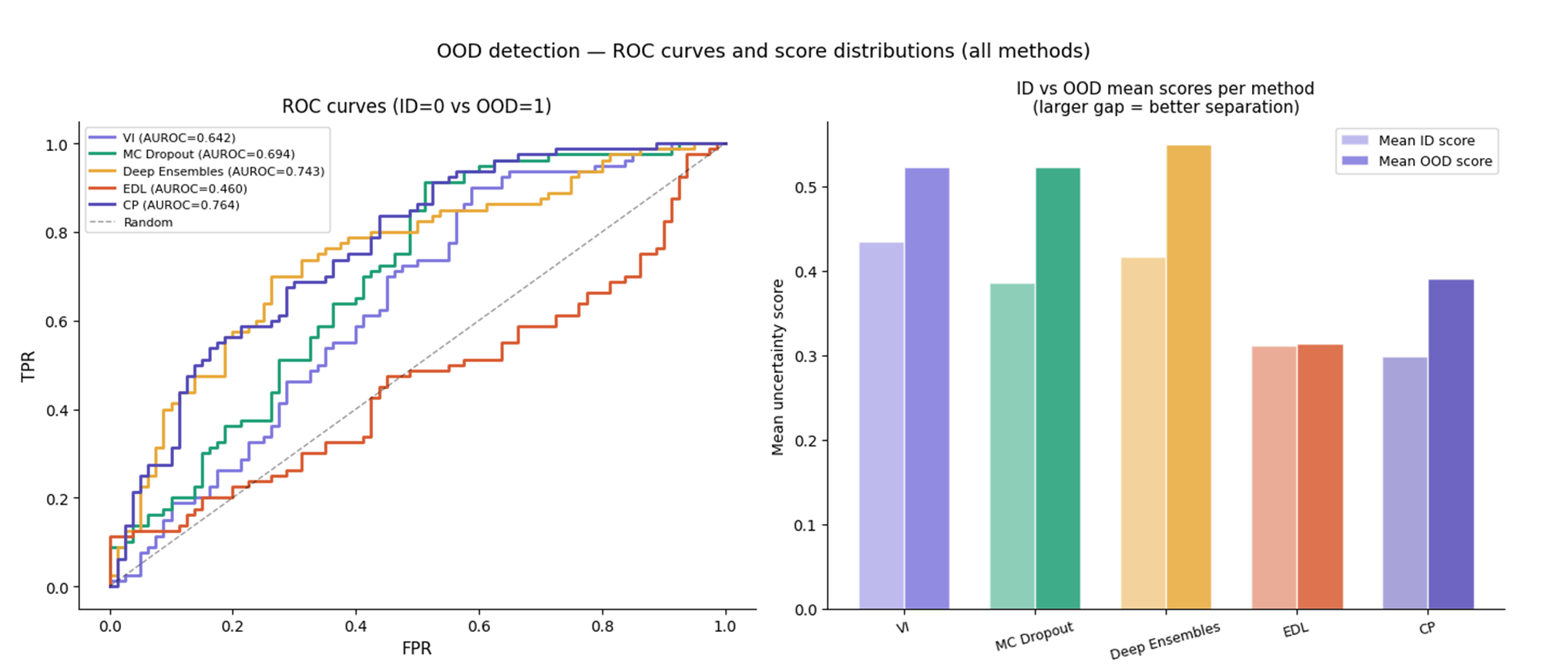}
\caption{OOD detection across all five methods. From left to right, receiver operating characteristic curves for separating in-distribution from OOD inputs using each method's uncertainty score; and mean uncertainty assigned to each group, where a larger gap indicates better separation.}
\label{fig:f4}
\end{figure*}

\begin{figure*}[t]
\centering
\includegraphics[width=\textwidth]{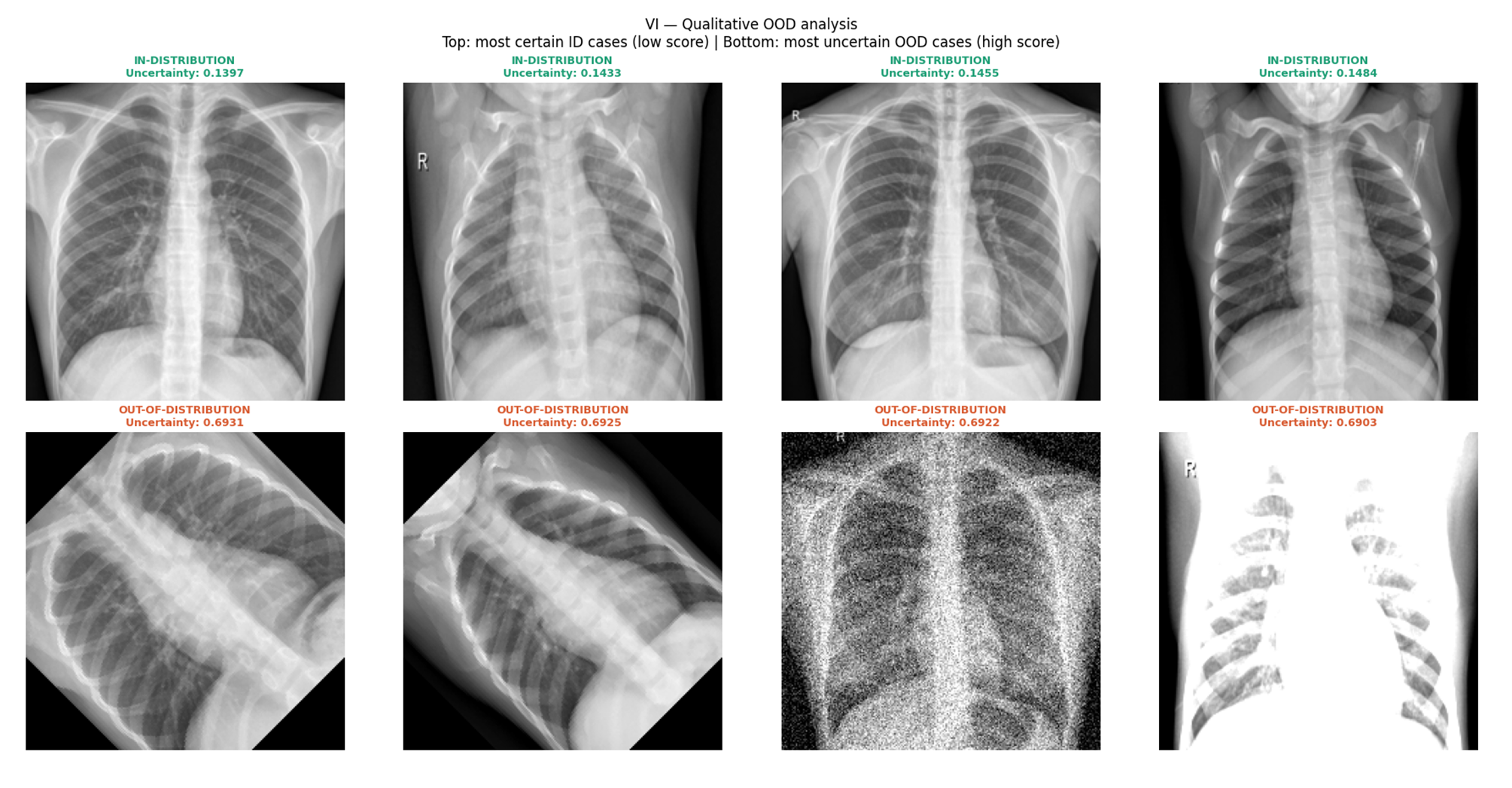}
\caption{Qualitative OOD analysis. Top row: in-distribution radiographs receiving the lowest uncertainty. Bottom row: OOD inputs, produced by rotation, noise, and intensity corruption, receiving the highest.}
\label{fig:f5}
\end{figure*}

\begin{table*}[t]
\centering
\small
\caption{Paired comparison across twenty language models. Chance accuracy is 0.25; an AUROC of 0.5 indicates confidence carrying no information about correctness.}
\label{tab:t3}
\setlength{\tabcolsep}{4pt}
\begin{tabular*}{\textwidth}{@{\extracolsep{\fill}}lccccccc@{}}
\toprule
\textbf{Metric} & \textbf{n} & \textbf{No context} & \textbf{Tutorial} & \textbf{Mean diff.} & \textbf{Median diff.} & \textbf{P} & \textbf{P (Holm)} \\
\midrule
Accuracy & 20 & 0.680 & 0.742 & +0.062 & +0.070 & 0.00016 & 0.00032 \\
AUROC & 20 & 0.725 & 0.788 & +0.063 & +0.069 & 0.00026 & 0.00032 \\
\bottomrule
\end{tabular*}
\end{table*}

\begin{figure*}[t]
\centering
\includegraphics[width=\textwidth]{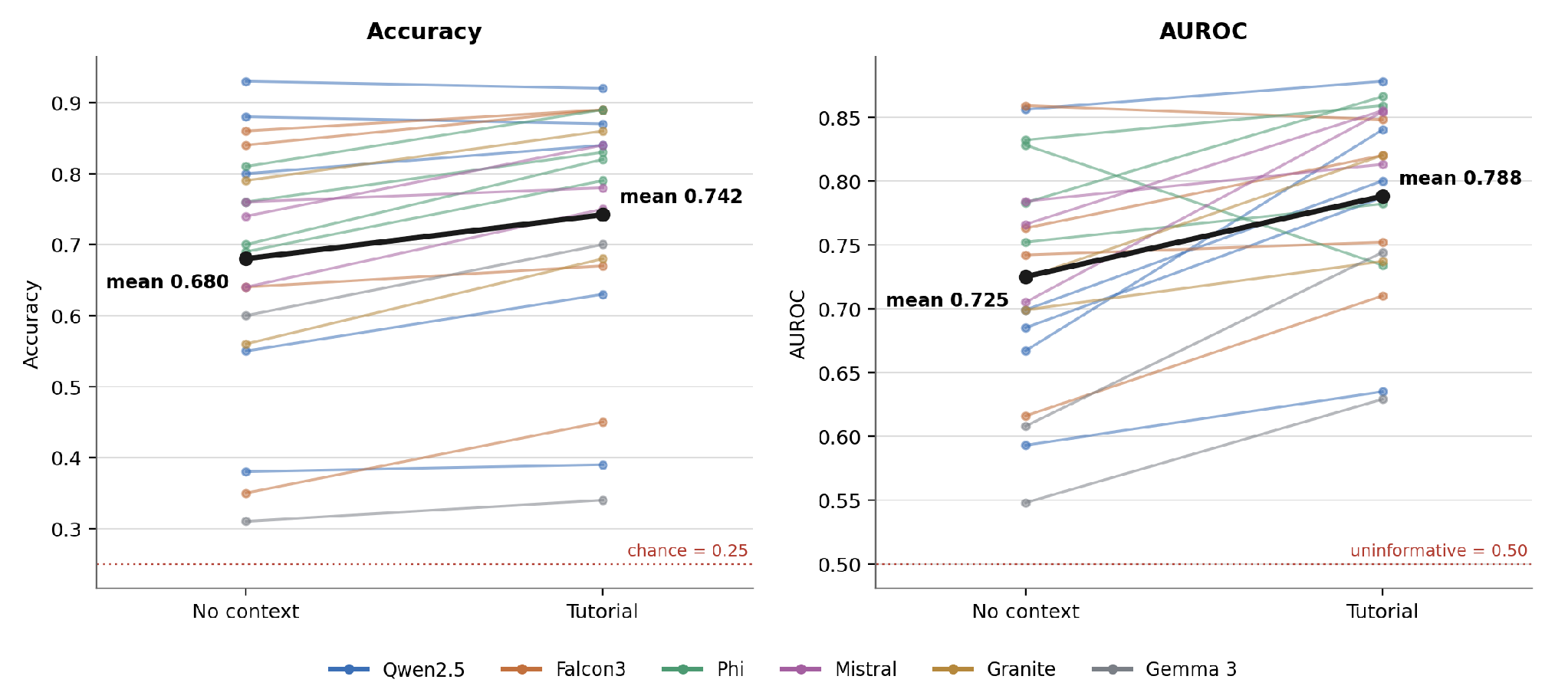}
\caption{Per-model paired differences between the no-context and tutorial conditions, colored by model family. Each thin line is one model; the heavy black line is the mean. Left: Accuracy. Right: AUROC.}
\label{fig:f6}
\end{figure*}

\begin{table*}[t]
\centering
\footnotesize
\caption{Per-model results. Acc accuracy, ctx context, AUROC area under the receiver operating characteristic curve.}
\label{tab:t4}
\setlength{\tabcolsep}{4pt}
\begin{tabular*}{\textwidth}{@{\extracolsep{\fill}}lccccccc@{}}
\toprule
\textbf{Model} & \textbf{Family} & \textbf{Acc. no ctx} & \textbf{Acc. tutorial} & \textbf{$\Delta$ Acc.} & \textbf{AUROC no ctx} & \textbf{AUROC tut.} & \textbf{$\Delta$ AUROC} \\
\midrule
Qwen2.5-0.5B & Qwen2.5 & 0.38 & 0.39 & +0.01 & 0.593 & 0.635 & +0.042 \\
Qwen2.5-1.5B & Qwen2.5 & 0.55 & 0.63 & +0.08 & 0.685 & 0.787 & +0.102 \\
Qwen2.5-3B & Qwen2.5 & 0.80 & 0.84 & +0.04 & 0.699 & 0.800 & +0.101 \\
Qwen2.5-7B & Qwen2.5 & 0.88 & 0.87 & $-$0.01 & 0.856 & 0.878 & +0.022 \\
Qwen2.5-14B & Qwen2.5 & 0.93 & 0.92 & $-$0.01 & 0.667 & 0.840 & +0.172 \\
Falcon3-1B & Falcon3 & 0.35 & 0.45 & +0.10 & 0.616 & 0.710 & +0.094 \\
Falcon3-3B & Falcon3 & 0.64 & 0.67 & +0.03 & 0.742 & 0.752 & +0.010 \\
Falcon3-7B & Falcon3 & 0.84 & 0.89 & +0.05 & 0.763 & 0.820 & +0.057 \\
Falcon3-10B & Falcon3 & 0.86 & 0.89 & +0.03 & 0.859 & 0.848 & $-$0.011 \\
Phi-3-mini & Phi & 0.69 & 0.79 & +0.10 & 0.783 & 0.866 & +0.083 \\
Phi-3.5-mini & Phi & 0.70 & 0.82 & +0.12 & 0.752 & 0.782 & +0.030 \\
Phi-4-mini & Phi & 0.76 & 0.83 & +0.07 & 0.828 & 0.734 & $-$0.094 \\
Phi-3-medium & Phi & 0.81 & 0.89 & +0.08 & 0.832 & 0.859 & +0.027 \\
Mistral-7B-v0.3 & Mistral & 0.64 & 0.75 & +0.11 & 0.784 & 0.813 & +0.030 \\
Ministral-8B & Mistral & 0.74 & 0.84 & +0.10 & 0.766 & 0.855 & +0.089 \\
Mistral-Nemo & Mistral & 0.76 & 0.78 & +0.02 & 0.705 & 0.854 & +0.150 \\
Granite-3.1-2B & Granite & 0.56 & 0.68 & +0.12 & 0.724 & 0.820 & +0.096 \\
Granite-3.1-8B & Granite & 0.79 & 0.86 & +0.07 & 0.699 & 0.737 & +0.038 \\
gemma-3-1b & Gemma 3 & 0.31 & 0.34 & +0.03 & 0.548 & 0.629 & +0.081 \\
gemma-3-4b & Gemma 3 & 0.60 & 0.70 & +0.10 & 0.608 & 0.744 & +0.137 \\
\bottomrule
\end{tabular*}
\end{table*}

\subsection{Tutorial Content and Representative Outputs}

Part 1 builds the clinical argument before introducing any mathematics. It separates the two ways a clinical model fails, confident error and indiscriminate caution, and argues that the first is the more dangerous because it is invisible. It then develops the aleatoric--epistemic distinction that organizes everything afterwards, and closes by observing that clinicians already reason in these terms, reporting a likelihood, a confidence, a differential, and a recommended action rather than a bare label.

Part 2 opens with the Bayesian view of a neural network, treating weights as distributions rather than fixed values, and works through the two approximations that make this tractable in practice, VI and MC dropout. Deep Ensembles arrive by a different route, requiring no special layers or loss functions: the same architecture is trained several times from different random initializations, and the disagreement among members is the uncertainty signal. EDL and CP follow, each taking its own path to the same goal. Every method is instrumented identically, so that the resulting quantities can be compared directly. Figure~\ref{fig:f1} illustrates the resulting uncertainty decomposition for a model, showing total predictive entropy alongside its epistemic and aleatoric components.

Part 3 addresses whether these quantities can be trusted. Calibration asks whether stated confidence matches observed frequency: among cases assigned 80\% confidence, approximately 80\% should be correct. The session measures this with reliability diagrams and expected calibration error (ECE), and applies temperature scaling as a post-hoc correction (Figure~\ref{fig:f2}).

Risk--coverage analysis addresses the operational question of when a model should decline to predict. Ordering cases by uncertainty and progressively abstaining on the least certain traces out a curve of error rate against the fraction of cases retained; the area under that curve summarizes the quality of the uncertainty ranking independently of any particular threshold (Figure~\ref{fig:f3}). This connects an abstract quantity to a deployment decision: what accuracy is achievable if a given share of cases is referred for human review.

OOD detection asks whether an input belongs to the training distribution at all. The session constructs corrupted and transformed radiographs as a controlled OOD set and evaluates how well each method's uncertainty score separates them from held-out in-distribution cases (Figure~\ref{fig:f4}). Figure~\ref{fig:f5} shows the qualitative counterpart, which is the form in which such behavior would be encountered in practice.

Part 4 compares the methods across the dimensions that determine which to use in practice, namely computational cost at training and inference, intrusiveness to an existing pipeline, quality of the resulting estimates, and the strength of any guarantee, and sets out the open problems.

\subsection{LLM Evaluation}

Accuracy improved in 18 of 20 models, increasing from a mean of 0.680 without tutorial context to 0.742 with retrieved tutorial context (mean difference, +0.062; median difference, +0.070; Holm-adjusted P = 0.00032). AUROC likewise improved in 18 of 20 models, increasing from 0.725 to 0.788 (mean difference, +0.063; median difference, +0.069; Holm-adjusted P = 0.00032) (Table~\ref{tab:t3}, Figure~\ref{fig:f6}).

Accuracy without context ranged from 0.31 (gemma-3-1b) to 0.93 (Qwen2.5-14B), and with context from 0.34 to 0.92. Per-model accuracy differences ranged from $-$0.01 to +0.12. The two models that did not improve were Qwen2.5-7B and Qwen2.5-14B, which scored 0.88 and 0.93 respectively without context and each answered one fewer item correctly with it. The largest accuracy gains, of +0.12, were recorded by Phi-3.5-mini and Granite-3.1-2B, both of which scored between 0.56 and 0.70 without context. AUROC without context ranged from 0.548 (gemma-3-1b) to 0.859 (Falcon3-10B), and with context from 0.629 to 0.878. Per-model AUROC differences ranged from $-$0.094 to +0.172. The largest gain was recorded by Qwen2.5-14B (0.667 to 0.840). Three models declined: Phi-4-mini ($-$0.094), Falcon3-10B ($-$0.011), and, marginally, no others, with Falcon3-3B recording the smallest positive change (+0.010). Phi-4-mini and Falcon3-10B both gained on accuracy over the same comparison (+0.07 and +0.03 respectively). All six model families improved in the mean on both metrics. The two metrics did not move together within models. Qwen2.5-14B recorded the largest AUROC gain (+0.172) alongside an accuracy change of $-$0.01, and Mistral-Nemo recorded +0.150 on AUROC against +0.02 on accuracy. Phi-3.5-mini recorded +0.12 on accuracy against +0.030 on AUROC. Per-model results are given in Table~\ref{tab:t4}.

\section{Discussion}

\subsection{An accessible and reproducible learning path for UQ in medical imaging}

This tutorial was developed to provide a structured, practical introduction to UQ for researchers and practitioners working with medical imaging models. Its primary contribution is not the introduction of new UQ methods, but the organization of established methods into an accessible, hands-on learning path that connects clinical motivation, methodological foundations, implementation, and evaluation. The tutorial moves from the reasons UQ is needed in medical imaging through several major methodological families and concludes with methods for assessing whether the resulting UQ estimates are informative and reliable. This final emphasis is important because obtaining an uncertainty measure does not by itself establish that the measure is useful. A tutorial that focuses only on producing uncertainty estimates risks treating the output of a UQ method as inherently trustworthy, whereas practical use requires consideration of calibration, selective prediction, distribution shift, and other aspects of reliability.

The tutorial also deliberately prioritizes accessibility and reproducibility. The implementation sessions use a common medical imaging task and model backbone, allowing readers to compare different UQ approaches within a consistent experimental setting rather than learning each method through a different dataset or architecture. Computational requirements were constrained so that the notebooks can be executed using freely available GPU resources. This design lowers the barrier to experimentation for readers who may not have access to institutional computing infrastructure and allows the conceptual material to be connected directly to executable implementations.

The scope of the tutorial reflects a deliberate balance between breadth and depth. It introduces several major families of UQ while providing enough theoretical background to motivate their implementation and sufficient practical material for readers to experiment with them. It is therefore intended as an entry point rather than a comprehensive treatment of any individual method. The practical component is currently centered on binary classification of chest radiographs, providing a consistent setting in which the different approaches can be implemented and compared. Although many of the underlying concepts extend to other medical imaging tasks, their practical formulation does not transfer directly. Segmentation, for example, raises questions about uncertainty at the pixel and anatomical-structure levels, while regression requires approaches to uncertainty over continuous outcomes. Extending the tutorial to these settings would broaden its applicability while preserving its implementation-oriented structure.

\subsection{Retrieved tutorial content improves LLM performance on independently constructed UQ questions}

The LLM evaluation supports a deliberately narrow claim. The tutorial contains information relevant to questions derived from the primary literature in UQ, and that information can be retrieved and used by LLMs to improve their responses. Because the questions were constructed from the primary methodological literature rather than from the tutorial itself, the evaluation does not simply test whether the models can reproduce information that was directly used to construct the question bank. Instead, it provides a test of whether the tutorial contains relevant knowledge that can be located and applied to independently constructed questions.

The pattern of results provides additional insight into where retrieved educational material may be most useful. Accuracy gains were greatest among models with intermediate baseline performance, whereas models at the upper end had limited room for improvement and models at the lower end showed more limited gains. Several factors could contribute to this pattern, including differences in baseline knowledge and in the ability to make effective use of retrieved context. The present study was not designed to distinguish between these explanations, but the results suggest that the benefit of external domain-specific context is not uniform across models.

The results also illustrate the value of evaluating more than a single performance measure. Although accuracy and AUROC capture different properties of model behavior, they do not necessarily change in parallel. The strongest models had limited room for improvement in accuracy, yet some still showed meaningful changes in confidence-based performance. This suggests that providing relevant external knowledge can affect aspects of model behavior that would not be captured by answer correctness alone. In this respect, the LLM evaluation mirrors a central principle of the tutorial itself: predictive performance and the reliability of confidence provide complementary information.

\subsection{LLM-based evaluation provides a scalable measure of technical content coverage}

More broadly, the experiment demonstrates a potential framework for evaluating technical educational resources at scale. Traditional evaluation of educational material requires human learners and can be time-consuming, particularly when the goal is to assess a large collection of technical concepts. LLM-based evaluation cannot replace such studies, but it can provide a complementary content-coverage assessment. In this setting, the approach is particularly useful because the question bank was constructed independently from the tutorial and the same evaluation set was used across all models. Retrieval was also performed once and cached, ensuring that differences in retrieval did not contribute to differences between models. Together, these design choices provide a reproducible framework for examining whether a technical resource can serve as an effective external knowledge source for contemporary LLMs.

\subsection{Limitations}

Several limitations should be considered when interpreting these findings. First, the tutorial is intended as an entry point rather than a comprehensive treatment of UQ. Its breadth requires a balance between theoretical depth and practical accessibility, and readers seeking to develop new UQ methodology will need to engage with the primary literature. The practical component is currently restricted to binary classification of chest radiographs, and although the underlying concepts are broader, the implementations do not directly address the distinct challenges posed by segmentation, regression, registration, reconstruction, or other imaging tasks.

Second, the LLM evaluation measures knowledge recognition and application in a multiple-choice setting rather than learning in the educational sense. Selecting the correct answer does not establish that a model can explain the underlying concept, reason through a novel clinical scenario, or retain the material over time. More importantly, LLM performance cannot be interpreted as a substitute for evidence from human learners. Demonstrating that the tutorial improves human understanding would require controlled studies involving appropriate learner populations and pre- and post-intervention assessment.

Third, the evaluation reflects both the quality of the tutorial and the effectiveness of the retrieval pipeline. A relevant concept cannot improve an answer if the corresponding material is not retrieved. Consequently, the observed effect represents the combined performance of the tutorial, its representation and chunking, the embedding model, and the retrieval strategy rather than the intrinsic quality of the tutorial alone. Future work could examine how alternative chunking strategies, embedding models, retrieval methods, and numbers of retrieved passages affect the results.

Finally, the question bank was generated by Claude Opus 5 from the primary literature. Although the generation prompt imposed explicit constraints on question construction, distractor quality, option length, and coverage, language-model-generated questions may still contain ambiguity or other forms of bias. Independent expert review of the questions and answer keys, as well as larger and independently constructed question banks, would strengthen future evaluations. The multiple-choice format also limits the range of abilities that can be assessed.

\subsection{Future directions}

Several extensions follow naturally from this work. Future versions of the tutorial could expand beyond classification to segmentation, regression, and other medical imaging tasks while incorporating additional UQ methods and more advanced clinical decision-making scenarios. The evaluation framework could likewise be extended to larger question banks, free-response questions, case-based assessments, and comparisons with alternative educational resources, including other massive open online courses. Varying the retrieval strategy would also allow the contribution of content quality to be separated more clearly from the contribution of retrieval quality.

Most importantly, future studies should evaluate the tutorial with human learners, including medical students, radiologists, and medical AI researchers. Pre- and post-intervention assessments could determine whether the tutorial improves conceptual understanding, implementation ability, and the interpretation of uncertainty in clinical scenarios. Such studies would complement the present LLM-based content-coverage evaluation by addressing the central question that the current study cannot answer, namely whether the tutorial improves human learning.

\subsection{Conclusion}

This work presents an open, implementation-oriented tutorial designed to provide a practical entry point to UQ for medical imaging. By connecting clinical motivation with major UQ methods, executable implementations, and evaluation of uncertainty reliability, the tutorial provides a structured path from understanding why uncertainty matters to assessing whether it can be trusted. Its consistent computational framework and use of freely available GPU resources are intended to make this material accessible to researchers and practitioners without specialized computing infrastructure.

We additionally demonstrate an approach for evaluating such a resource as an external knowledge source for LLMs. Across 20 instruction-tuned models, retrieval of the tutorial improved performance on an independently constructed set of UQ questions, with effects that varied across models. These findings support the use of LLM-based evaluation as a complementary measure of whether technical educational material contains retrievable and usable domain knowledge, while not providing evidence of human educational effectiveness. Future human-learning studies will be needed to determine whether the tutorial translates this accessible knowledge resource into improved understanding and practice of UQ in medical imaging.

\section*{Data Availability}

All tutorial notebooks are publicly available at \url{https://github.com/benyamin-gheiji/Uncertainty-Quantification-Medical-Imaging-Analysis/}.

\section*{Acknowledgements}

The authors thank Dr. Chen Chen for her valuable guidance and thoughtful input during the development of this tutorial.

\end{document}